\documentclass[epj]{svjour}
\usepackage[UKenglish]{babel}
\usepackage[utf8x]{inputenc}

\usepackage[square,numbers,comma,sort&compress]{natbib}
\usepackage{subfigure}% Support for small, `sub' figures and tables
\usepackage{amsmath}
\usepackage{amssymb}
\usepackage[pdftex]{graphicx}
\usepackage[dvipsnames]{xcolor}
\usepackage[colorlinks=true, allcolors=blue]{hyperref}
\usepackage{enumerate}
\usepackage{wasysym}
\usepackage{tabularx}
\usepackage[sc]{mathpazo}
\usepackage{orcidlink}

\begin{document}
\sloppy		% without this option, text that latex can't figure out how to hyphenate sometimes runs over the designated column width

\title{Quantum Battles in Attoscience - Tunnelling}
%\titlerunning{Tunnelling Battle}
% \subtitle{Do you have a subtitle?\\ If so, write it here}
\author{Cornelia Hofmann\inst{1}\orcidlink{0000-0001-7183-5564}, Alexander Bray\inst{2}\orcidlink{0000-0002-3780-3312}, Werner Koch\inst{3}, Hongcheng Ni\inst{4,5}\orcidlink{0000-0003-4924-0921} \and Nikolay I. Shvetsov-Shilovski\inst{6}\orcidlink{0000-0002-8825-9302}% etc
% \thanks is optional - remove next line if not needed
%\thanks{\emph{Present address:} Insert the address here if needed}%
}                     % Do not remove
\authorrunning{C. Hofmann, A. Bray, W. Koch, H. Ni \and N. I. Shvetsov-Shilovski}
%
% \offprints{}          % Insert a name (to order prints) or remove this line
\mail{c.hofmann@ucl.ac.uk}
\institute{Department of Physics \& Astronomy, University College London, Gower Street, London WC1E 6BT, United Kingdom
	 \and Research School of Physics, The Australian National University, Canberra, Australian Capital Territory 0200, Australia 
	 \and Weizmann Institute of Science, Rehovot, Israel
	 \and State Key Laboratory of Precision Spectroscopy, East China Normal University, Shanghai 200241, China
	 \and Institute for Theoretical Physics, Vienna University of Technology, A-1040 Vienna, Austria, European Union
	 \and Institut für Theoretische Physik, Leibniz Universität Hannover, D-30167 Hannover, Germany, European Union
  }
\date{Received: date / Revised version: date}
% The correct dates will be entered by Springer
%
\abstract{
What is the nature of tunnelling? This yet unanswered question is as pertinent today as it was at the dawn of quantum mechanics. This article presents a cross section of current perspectives on the interpretation, computational modelling, and numerical investigation of tunnelling processes in attosecond physics as debated in the Quantum Battles in Attoscience virtual workshop 2020.
} %end of abstract
\maketitle
\section{Introduction}
\label{intro}

%\begin{itemize}
%	\item Quantum tunnelling importance/relevance/\ldots
%	\item temporal resolution: debated theoretically for a long time, experimentally since XXXX,
%	\item new high interested in the temporal resolution due to its importance in many attoscience techniques
%	\item comment on conference/battle structure		
%	\begin{itemize}
%		\item "Battle" session during the "Quantum Battles in  Attoscience" 2020 virtual workshop
%		\item designed to be an open debate among several combatants who have published on the contentious topic of the mathematical description, understanding and possible temporal resolution of quantum tunnelling. 
%		\item In addition to questioning each other, the moderator also posed live audience questions to the combatants. 
%		\item This resulted in a highly interactive and lively debate \cite{PhysicsWorldControversy2020} on an issue which has seen a rather controversial back-and-forth in the literature \rot{[some examples...]}.
%		\item This perspective article offers a summary of that debate. 
%	\end{itemize}
%\end{itemize}

The discovery of the quantum tunnelling phenomenon almost 100 years ago has not only opened up many new avenues and applications.
It has also kept quantum physics researchers busy since then, trying to define the temporal resolution of the process \cite{Hauge1989,Landauer1994}. 
Early experiments were focused on photons tunnelling through potential barriers, such as Ref.~\cite{Steinberg1993} for example. 
But with the advent of attosecond science \cite{Corkum2007} the question "\emph{Does tunnelling take time, and if yes, how much?}" has gained a lot of new interest, since electron dynamics often include quantum tunnelling portions, be that in biological processes such as photosynthesis \cite{Filho2019} or charge transport in semiconductors \cite{Teuscher2017a}, tunnelling ionisation as the first step for high-order harmonic generation (HHG) spectroscopy \cite{Bruner2015}, photoelectron holography \cite{Huismans2011a}, laser induced electron diffraction (LIED) \cite{Meckel2008} or many more. 

The temporal resolution of quantum tunnelling is still heavily debated \cite{Hofmann2019,Kheifets2020,SatyaSainadh2020,Ramos2019} and thus presented an interesting topic for a debate at the Quantum Battles in Attoscience virtual workshop 2020 \cite{Hauge1997}. The aim of the \emph{Battle} sessions was "an open debate on a contentious topic involving several early career researchers ('combatants') and the entire audience of attendees" \cite{QuantumBattlesWebsite}. To that effect, the combatants prepared a scaffolding structure of the debate on "tunnelling", defining three main topics: a) Physical observables and typical experiments (presented in section \ref{Sec:Physical} of this article), b) Nature of Tunnelling (see section \ref{Sec:Philosophical}), and c) Theoretical approaches to quantum tunnelling time (in section \ref{Sec:Numerical}). Each topic was introduced with an overview presentation, followed by a free debate among all combatants, moderated by Prof. Jonathan Tennyson, UCL, and included both questions among the combatants as well as live audience questions. 
The result was a highly interactive and lively debate \cite{PhysicsWorldControversy2020}.

This perspective article offers a text-form of the live debate \cite{Battle1YouTube}, supplemented with additional references and explanations.

\section{Physical observables and typical experiments}
\label{Sec:Physical}

The guiding questions for this first topic are: 

\begin{quote}
	\emph{What are physical observables, typical measurements, and what are the characteristic physical systems under investigation? 
		\\What other aspects of these particular systems influence the interpretation of tunnelling time studies?}
\end{quote}

\subsection{Overview:}

When it comes to experiments investigating the temporal resolution of a quantum tunnelling particle, there are typically two kinds of experiments: 
a) Bose-Einstein-Condensates (BEC) of atoms trapped in optical lattices, with various manipulations on them to measure tunnelling from one lattice site to the next \cite{Fortun2016,Ramos2019}. 
Since the particles in question are entire atoms, their temporal resolution for the dynamics is in the range of microseconds. 
And b) attosecond angular streaking (also known as attoclock) type \cite{Landsman2013a,Camus2017,Sainadh2019} experiments, a technique developed in strong-field attosecond physics, where electrons tunnel ionise from a bound state through the potential barrier which is created by the interaction of the strong laser field with the binding Coulomb potential of atoms. 
These are on the attosecond regime since electrons are tunnelling, and the main focus of the here following debate. 

\begin{figure}[h]
	\centering
	\includegraphics[width=0.8\columnwidth]{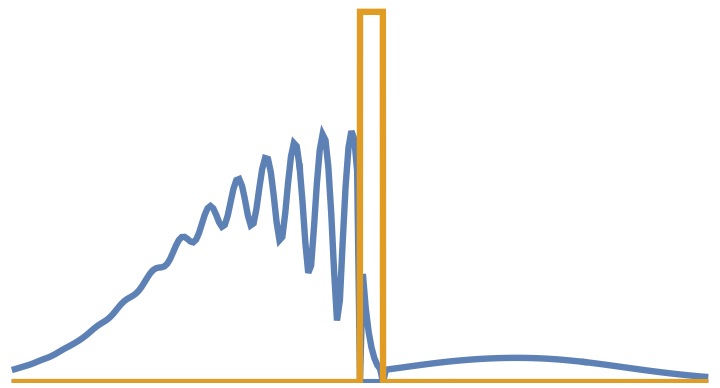}
	\caption{Idealised sketch of a wave packet hitting and partially tunnelling through a potential barrier.}
	\label{fig:wavepacketframe43cropped}
\end{figure}

On a fundamental level, what we are interested in is the temporal resolution of a wave packet hitting a potential barrier, and then a part of that wave packet tunnelling through, such as schematically illustrated in figure \ref{fig:wavepacketframe43cropped}. 
However, this exactly creates several challenges in trying to time this process compared to other timings of wave packets, such as for example group delay in photonics. 
The peak of the wave packet is not conserved, since the incoming (or bound state) wave packet is split into a reflected and a transmitted part. 
The potential barrier essentially acts as an energy-dependent filter, such that the spectra of the two resulting wave packets are significantly different \cite{Hofmann2019}.
A wave packet also always corresponds to a probability distribution, and in consequence it is difficult to define a clear starting and ending (or entrance and exit) point, more on that in section \ref{Sec:Philosophical}.
Furthermore, in strong-field attoscience scenarios we are tunnel ionising from a bound state, where of course parts of the wave function even in its field-free ground state always are "under" the barrier, without any tunnelling occurring.
Additionally, approaches such as Wigner-like, scattering and resonance phase times \cite{Isinger2017} which are commonly applied to single-photon ionisation \cite{Gallmann2017} are not applicable either, again because of the chirped propagation of the electron wave packet and the energy filtering of the potential barrier. 
While we are on the topic of potential barriers, it is also worthwhile noting that this classical picture of the potential barrier only emerges if the laser field is treated in the length gauge \cite{Reiss2008,Reiss2008erratum,Reiss2014,Reiss2021}.

The physical observable for measurements (and calculations often, too) of strong-field tunnel ionisation are momentum distributions of photoelectrons \cite{Landsman2014b,Camus2017,Sainadh2019} or momenta of atoms \cite{Fortun2016,Ramos2019}. 
Momentum is of course a standard quantum mechanical observable corresponding to a unitary operator, whereas time itself is a parameter of the Schrödinger equation and thus not an observable as such.
Therefore, a relation between measured (or calculated) momenta and the timing of the tunnelling process needs to be established through theoretical understanding of the quantum tunnelling process.

In the experiment by Fortun and co-workers a rubidium BEC is oscillating in an optical lattice. 
In a pump-probe-type approach, the lattice is turned off at different intervals after the initiation of the oscillation and the instantaneous momentum of the atoms carried them flying towards a position-sensitive detector. 
The tunnelled wave packets appeared delayed with respect to the reflected wave packets \cite{Fortun2016}.  
In the experiment by Ramos and co-workers, a quantum simulation of the Larmor clock \cite{Buttiker1983,Baz1967,Rybachenko1967}, one of the well known theoretical approaches to predicting the tunnelling time \cite{Landsman2015}, was realised causing precession of the spin of the rubidium atoms while traversing a potential barrier. 
This spin precession was then mapped onto different states according to the angle of rotation and separated by a Stern--Gerlach measurement \cite{Ramos2019}. 

In attoscience experiments utilising the attoclock method \cite{Eckle2008,Eckle2008a}, the rotation of the nearly circularly polarised vector potential $\mathbf{A}$ mimics the hand of a clock.
The path of a photoelectron after tunnel ionisation is dominated by the interaction with the laser field \cite{Hofmann2019CTMCvideo}, and thus neglecting all other corrections and perturbations, the final asymptotic momentum $\mathbf{p}_f$ is determined by the vector potential at the time when it first exits the potential barrier and enters the continuum $t_0$, through the conservation of canonical momentum 
\begin{equation}
\mathbf{p}_f = \mathbf{p}_0 - \mathbf{A}(t_0),
\end{equation} 
where $\mathbf{p}_0$ denotes a possible initial momentum.
Hence, the final momentum angle acts as a clock for the exit moment in time. 
However, this angle to time mapping is subject to several corrections, some easy to describe and include in calculations, others more elusive to quantify and thus the topic of ongoing research. 
A non-exhaustive list of corrections, approximations, and other issues include: 
the Coulomb force of the parent ion induces an angular shift \cite{Torlina2012,Sainadh2019,Bray2018b,Kheifets2020};
the ellipticity, pulse envelope, pulse duration, and carrier-envelope-offset phase are wave form parameters which affect the photoelectron trajectories; 
and depletion mixes in with pulse duration and the intensity of the applied field \cite{Torlina2015,Ni2018}
for topics which mostly have been dealt with in great detail and comparable results;
the experiment does not have access to any "start" signal of the tunnelling process, only the exit point \cite{Landsman2014b,Hofmann2019}; 
non-adiabatic effects influence the ionisation rate, energy at tunnel exit, initial momentum $\mathbf{p}_0$ distribution, and the location of said tunnel exit itself \cite{Ivanov2014,Klaiber2015,Ni2016,Ni2018,Ni2018a}; 
multi-electron-effects are ignored in most calculations \cite{Lezius2001,Pfeiffer2012,Shvetsov-Shilovski2012,Emmanouilidou2015,Majety2017}; 
models including non-classical characteristics of the trajectory which can be compared against experimental data are still being developed \cite{Torlina2015,Camus2017,Shvetsov-Shilovski2019a}; 
and the orbital angular momentum of the bound state has an effect on the  strong-field ionisation \cite{Beiser2004,Barth2011,Herath2012,Liu2018}
for issues which are more elusive (although this categorisation is not definite). 

\begin{figure}[h]
	\centering
	\includegraphics[width=0.8\columnwidth]{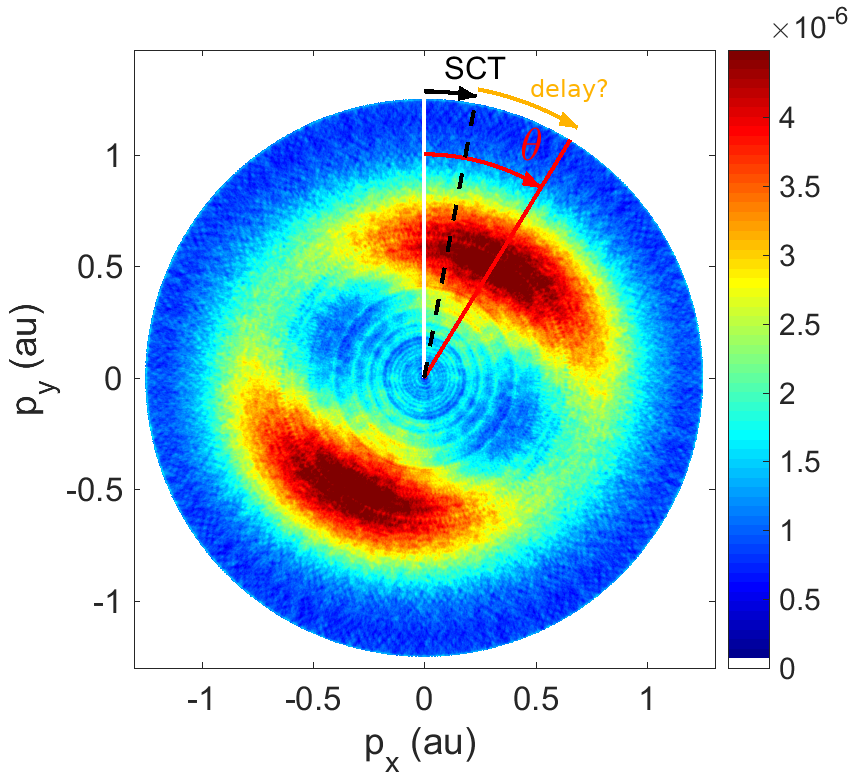}
	\caption{Illustration of photoelectron momentum distribution for ellipticity 0.87, clockwise helicity, projected to the plane of polarisation. Single Classical Trajectory (SCT) models assuming instantaneous tunnelling predict an angle offset away from the pure $-\mathbf{A}(t_{max})$, but the measured angle offset might be even larger than that. Adapted from \cite{Hofmann2019}.}
	\label{fig:vmidata2}
\end{figure}

There is still a lot of work necessary to properly disentangle the different contributions which lead to various angular shifts, sketched in figure \ref{fig:vmidata2} of the measured Photoelectron Momentum Distribution (PMD), until we can be sure of the remaining angle offset and it's relation to tunnelling time. 

\subsection{Debate:}
\begin{itemize}
	\item %short vs long pulses
	\begin{figure}[h]
		\centering
		\includegraphics[width=\columnwidth]{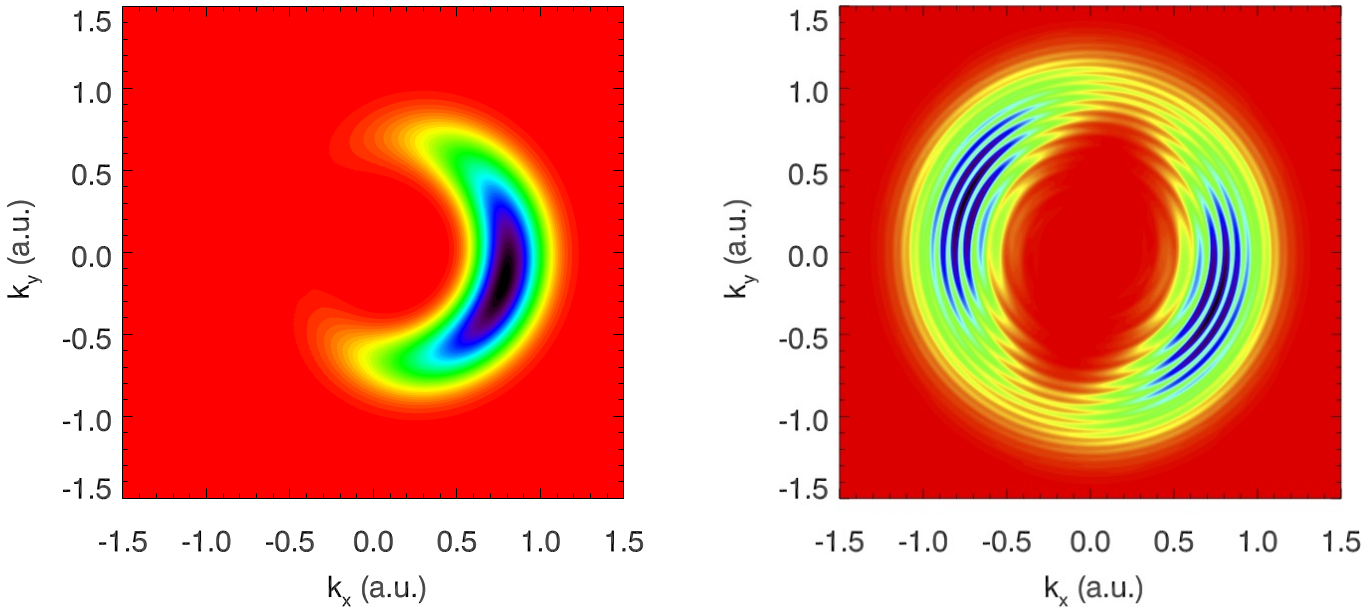}
		\caption{Time-dependent Schrödinger equation (TDSE) calculation of photoelectron momentum distributions for hydrogen ionisation. \textbf{Left}: idealised attoclock with a single cycle pulse and circular polarisation leads to a unique final momentum probability distribution peak. Pulse duration $\approx 1.6~ \mathrm{fs}$ FWHM, peak intensity $0.86 \times 10^{14}~\mathrm{W/cm^2}$, wavelength 800~nm, with clockwise helicity. \textbf{Right}: A multi-cycle pulse yields two main blobs with Above Threshold Ionisation (ATI) rings from the inter-cycle-interference. Pulse duration $\approx 6~\mathrm{fs}$ FWHM, peak intensity $1.5 \times 10^{14}~ \mathrm{W/cm^2}$, wavelength 770~nm, with clockwise helicity. Adapted from \cite{Kheifets2020,Bray2020}.
			%\rot{maybe flip one of them, so the rotation sense is the same for both of them?}
		}
		% Alex thesis caption: 
%		The short pulse is approximately 1.6 fs FWHM in intensity of peak 0.86 × 1014 W/cm2, 800 nm, ellipticity 1.0, and anti-clockwise helicity. Its corresponding distributions assume CEP stability. The long pulse is approximately 6 fs FHWM in intensity of peak 1.5 × 1014 W/cm2, 770 nm, ellipticity 0.85, and clockwise helicity. Its corresponding distributions are averaged over CEP
		
		\label{fig:alexconversationstartersshortlong}
	\end{figure}
	Figure \ref{fig:alexconversationstartersshortlong} exemplifies the pulse duration and wave form dependence, as well as an energy dependence between the different ATI rings in the long pulse case, which show different angular maxima \cite{Xie2017a,Eicke2019,Yuan2020}.
	This raises concerns about the validity of one single time (rather than a distribution of times) extracted typically from data, thus averaging over the energy dependence. 
	In attoclock experiments, the carrier-envelope-offset phase (CEP) was not stabilised \cite{Landsman2014b,Camus2017,Sainadh2019}. Additionally, the orientation of the polarisation ellipse in the lab frame was chosen such that the observable of interest (angular shift mostly parallel to the major axis of polarisation) was orthogonal to the direction with the biggest experimental noise (along the gas jet direction, thus chosen for the minor axis of polarisation) \cite{Doerner2000}. 
	Both of these effects wash out ATI interference. 
	For the CEP influence in particular, the interplay between ellipticity and pulse duration is critical. 
	For the largest field strength to be following the polarisation ellipse (desired in attoclock experiments \cite{Hofmann2019}) rather than the CEP \cite{Eckle2008,Eckle2008a}, the pulse envelope must be long enough relative to the ellipticity reducing the field strength within a quarter cycle. 
	Furthermore, ATI rings result from interference created by many laser optical cycles, highlighting the difficulty of defining a "single time", or even relative time intervals with respect to local maxima of the field strength or other possible references. 
	
	\item % distribution of the momenta, structure $\Rightarrow$ probabilistic distribution of times?
	Most often, tunnelling time calculations tend to use only a single peak point in the momentum distribution \cite{Landsman2014,Torlina2015,SatyaSainadh2020}. 
	However, based on this discussion it would seem more appropriate to extract the tunnelling time from the full momentum distribution, which contains much more information regarding the tunnelling process \cite{Landsman2015,Ni2018}.
	We note that such work has been carried out in a recent publication \cite{Ma2021}, which assesses the whole momentum distribution instead of just a single offset angle.
	
	\item %Question: how is the peak of the distribution determined precisely, i.e. maximum in 2D distribution is not the same as the maximum in the 1D angular distribution?
%	\begin{itemize}
%		\item relationship angle-time strictly linearly only works for circular polarisation. Elliptical geometry introduces another correction \cite{Hofmann2013}
%		\item polarisation plane cut vs projection of the propagation direction onto the polarisation plane: is there any dependence on that third momentum component? Any extra physics involved in this direction? For now, it seems as if this possible dependence should not mess with the time-delay analysis
%	\end{itemize}
	An audience question is brought in: How is the peak of the PMD determined precisely, since the maximum in a 2D distribution is not the same as the maximum in the 1D angular distribution?
	Of course the strictly linear angle-time relationship is only exact for circular polarisation.
	For any other polarisation, the elliptical geometry introduces corrections and needs to be taken into consideration \cite{Hofmann2013}.
	These effects as well as the influence of integrating over the radial component in the 2D distribution were double-checked against.  
	The resulting shifts in the extracted values were smaller than or of the order of the reported error bars for experimental data. 
	Nevertheless, it is important to keep in mind that different coordinate system transforms and peak angle extraction methods lead to significant shifts in the extracted angle and thus the interpreted delay time \cite{Eicke2018}.
%	\rot{(Comment: Are you sure? As far as I know, the articles of Manfred Lein says that the Jacobian of coordinate transform leads to a big change in the angle, For example: PhysRevA.97.031402(2018).)}
	Regarding the third component, the laser propagation direction, so far no significant difference has been found between a projection to or a cut along the polarisation plane.
	Of course this requires that no extra physics becomes important along this third component, for example the influence of the magnetic field must be negligible \cite{Reiss2014,Ludwig2014}. \\
	Of course, an energy-resolved angular distribution would avoid the integration over at least one of the components and thus make the peak search less dependent on geometry and coordinate choices.

%	\item Question: Wigner-Smith time delays depend on electronic potential landscapes\ldots \rot{I believe this questions actually was meant for the talk before rather than our battle discussion, or at least it would fit in there much better}
\end{itemize}

\section{Nature of tunnelling}
\label{Sec:Philosophical}

The guiding questions for the second topic are: 

\begin{quote}
	\emph{What is the nature of tunnelling at the classical/quantum intersection? \\
		What is the "beginning" and "end" of tunnelling, and how do we define it? \\
		What are classical or quantum trajectories?}
\end{quote}

\begin{figure}[h]
	\centering
	\includegraphics[width=\columnwidth]{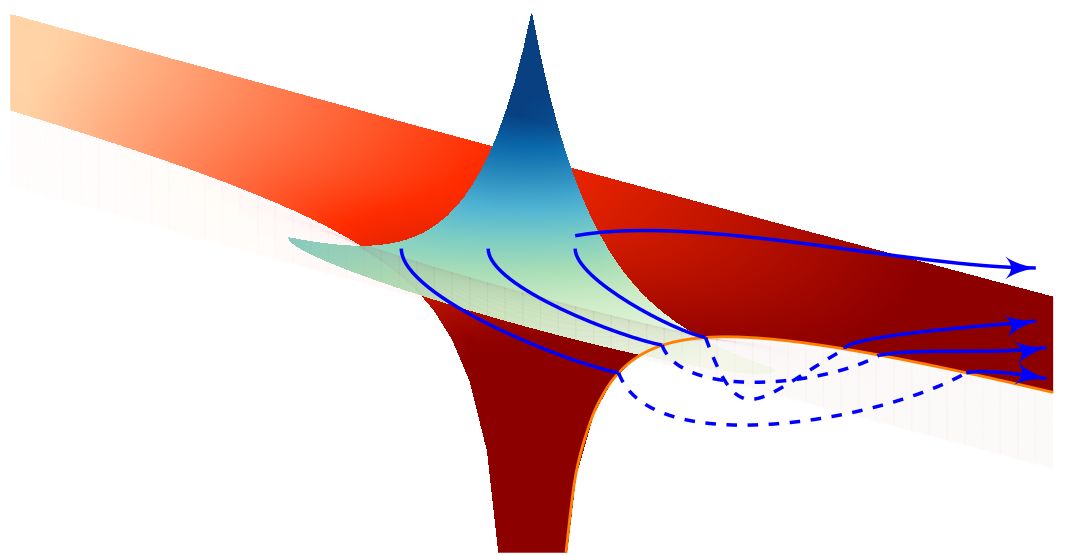}
	\caption{A central conundrum of quantum tunnelling: Wave functions tunnel naturally but have no clear tunnel entry or exit. Real valued trajectories allow for a clearly defined tunnel entry and exit criterion but can not tunnel without excursions into the complex plane. Which of the two perspectives is the better choice?}
	\label{fig:wernerintroimage}
\end{figure}

\subsection{Overview:}
%\begin{itemize}
%	\item Quantum domain: 
%	\begin{itemize}
%		\item tunnelling is a wave phenomenon 
%		\item $\Rightarrow$ where and when does it start? 
%		\item QM doesn't care about "tunnelling" (i.e. tunnelling is an intuitive but classical idea in our heads)
%	\end{itemize}
%	\item classical domain: 
%	\begin{itemize}
%		\item well defined in space-time 
%		\item $\Rightarrow$ can classical trajectories tunnel?
%	\end{itemize}
%		\item is a synthesis possible?
%		\begin{itemize}
%			\item waves transformed to quantum orbits, ensembles of trajectories?
%			\item which quantities best characterise the onset and end of tunnelling?
%		\end{itemize}
%\end{itemize}

Quantum tunnelling is a wave phenomenon, and the Time-dependent Schr\"{o}dinger equation (TDSE) is an equation for the probability amplitude wave (wave function). 
But this description makes it difficult to define where and when tunnelling exactly starts. 
Tunnelling itself is natural in quantum mechanics, it is only when we look at it from a classical perspective that there is a "forbidden" region in the potential barrier. 
In the classical domain, trajectories are well defined in space and time, but can they tunnel?
%On the other hand, 
The semiclassical models typically use classical trajectories to describe the motion of an electron after it has been released from an atom, usually by tunnelling ionisation.

Is a synthesis of these two worlds like the sketch in figure \ref{fig:wernerintroimage}, aiming to retain the quantum physics behaviour with the clarity of trajectories, possible?
% The question then whether the synthesis of these two completely different ways of the description, i.e., quantum and classical, is possible.  
It is clear that such a synthesis is not a simple task.  Indeed, in order to calculate the classical trajectory, i.e., to integrate Newton’s equation of motion, both starting point and the initial velocity are needed. However, Heisenberg’s uncertainty principle imposes a fundamental limit to the accuracy with which the values of the position and momentum, as well as of any other canonically conjugate variables, can be simultaneously determined.
Nevertheless, the application of the quasiprobability distribution allows to obtain information about both the position and momentum from the wave function. 
The most widely known examples of quasiprobability distributions are the Wigner function and Husimi distribution. 
We note that the Wigner function has already been used for description of strong-field processes, see, e.g., Refs.~\cite{Czirjak2000,Grafe2012,Chomet2019}. %Chirila2010,Yuan2012,Wu2013 not using Wigner function
However, to the best of our knowledge, the Wigner function has not yet been applied to the combination of the quantum and trajectory-based description in strong-field tunnel ionisation,
although a similar method has been proposed for the case of attosecond pulse single-photon ionisation with subsequent streaking of the photoelectron wave function \cite{Zimmermann2018}.
A recent and successful attempt of such combination was made in Ref.~\cite{Shvetsov-Shilovski2019a} using Gabor transform.

Doing so still begs the question, which quantity best characterises the onset of tunnelling?

\subsection{Debate:}
\begin{itemize}
	\item %Tunnelling is both quantum and classical: 
%	\begin{itemize}
%		\item quantum particles necessary, but barrier defines local properties significant for a classical system. So we need a combination of both: quantum tunnelling feature, but classical flavour of understanding/picture
%		\item but can we still look at the entire process in one single picture then? Can we rely on real-space trajectories? To include the tunnelling phenomenon, we need to include complex space and time trajectories, resulting in a quantum trajectory with clearly defined entry and exit to the barrier, as well as corresponding times. 
%		\item experiment can find (real) tunnel exit time. Hence everything happening after tunnelling should be described classically (classical trajectories) for experimental comparison. 
%		\item experimental comparison is done on the basis of the final momentum distribution, not the situation at the exit point, so purely quantum models can also be used for that purpose as long as they can predict a final momentum distribution, with certain assumptions built into the model (i.e. the zero-time prediction vs experimental prediction, then the difference is presumably due to a delay not accounted for in the model)
%	\end{itemize}
	Quantum particle description is necessary for tunnelling to occur in the first place, but the potential barrier defines local properties which are significant for classical systems. So we need a combination of both, quantum tunnelling feature with the classical flavour of understanding if we aim for any kind of temporal resolution of a tunnelling process. 
	The challenge is then to find one single picture for the entire process. 
	Instead of relying on real-space trajectories, including complex space and time enables the tunnelling phenomenon, resulting in a quantum trajectory with clearly defined entry and exit to the barrier \cite{Koch2020}, as well as corresponding times (more on this method in section \ref{Sec:Numerical}). 
	
	On the other hand, measurements can always only find real observables, thus fully complex calculations must find their way to the real axis somehow, where the propagation of a photoelectron wave packet is very well described by classical methods \cite{Ehrenfest1927}.
	But since the experimental observables typically are momenta, 
%	relied by position (and time-of-flight, for some) of the impact on a detector, 
	purely quantum models which operate in complex space and time can still be used for the purpose of comparison, as long as they can predict a final momentum distribution. 
	
	\item %reference point of field maximum (for the "starting time")
%	\begin{itemize}
%		\item missing half of the effect
%		\item the tunnelling process in strong-field ionisation is not necessarily a symmetric problem
%		\item field maximum as reference point is a chosen assumption, and studies trying to identify a physical starting point have found other values (typically before the maximum is reached)
%	\end{itemize}
	One huge assumption in experimental approaches based on the "attoclock" principle is the "starting time", relative to which the tunnelling delay is calculated. 
	This is typically chosen to be the maximum of the electric field, since that moment corresponds to the highest probability of tunnelling \cite{Hofmann2019}.
	However, this assumption might be missing out on half of the effect \cite{Klaiber2015}, and the tunnelling process in strong-field ionisation might be a symmetric problem relative to the (local) field maximum \cite{Ivanov2018}. 
	Publications which attempted to identify a physical starting point have found other values, typically before the maximum is reached \cite{Teeny2016,Teeny2016a,Ni2016,Klaiber2015}.
	
	\item %distinguishing tunnelling from over-the-barrier ionisation (OBI)
%	\begin{itemize}
%		\item in the experimental final momentum distribution it is not possible to a-posteriory separate these
%		\item in theoretical calculations based on trajectories these two processes can be differentiated. 
%		\item Quantum wave function calculations however would again not allow the disambiguation. 
%		\item non-adiabatic effects also play into these definitions, i.e. at what energy/distance a photoelectron can exit the potential barrier, or when the onset of OBI occurs
%		\item Backpropagation method \cite{Ni2016,Ni2018,Ni2018a} can easily determine tunnelled or OBI characteristic of any trajectory
%	\end{itemize}
	If we consider fully quantum models which describe both the tunnelling transition from bound to ionised state and the propagation afterwards in one, it becomes important to distinguish tunnelling from over-the-barrier (OBI) ionisation. 
	In experimental approaches it is generally not possible to \emph{a-posteriori} separate these two contributions to the total momentum distribution, and the same limitation is also true for numerical solutions of the TDSE \cite{Krainov2008}.
	
	However, in theoretical calculations based on trajectories, these two processes can easily be differentiated. 
	The semiclassical models naturally distinguish between the tunnelling through a potential barrier and the over-barrier-ionisation. Indeed, when the field strength is so high that the potential barrier formed by the laser field and the ionic potential is suppressed, it is impossible to find the starting point of the electron trajectory using field direction model (see, e.g., Refs.~\cite{Brabec1996,Chen2002,Delone1991,Lemell2012}) or the separation of the static tunnelling problem in parabolic coordinates \cite{Landau1965}. In this case it is usually assumed that the electron starts at the top of the suppressed potential barrier, and the difference between the ionisation potential and the energy at the top of the barrier $\Delta E=-I_p-V_{\text{max}}$
%	\rot{(Comment: I think the sign should be opposite? My convention used for this argument is $I_p>0$ and $V_\text{max}<0$. Are you using the same convention?)} 
	is transferred to the initial longitudinal velocity of the departing electron: 
		\begin{equation}
		\label{parvel}
		v_{0,\parallel}=\sqrt{2\Delta E}.
		\end{equation}
%	\rot{comment: we're all using different coordinate systems for major/minor axis \& propagation direction, so instead of labelling the longitudinal component as $z$, I think $\parallel$ is more general?}
	
	Non-adiabatic effects, i.e. effects beyond the quasistatic approximation which are due to the time-dependent changes in the strong field, also play into these definitions.
	For example, at which energy or distance can a photoelectron exit the potential barrier \cite{Mur2001,Ni2018a} or when does the onset of OBI occur?
	The so-called backpropagation method \cite{Ni2016,Ni2018,Ni2018a} is one hybrid approach which utilises the full quantum power of the TDSE for the tunnel ionisation but then retroactively adds the power of classical trajectories to also distinguish between OBI and tunnelling (more on this method in section \ref{Sec:Numerical}).

	\item %Question: localised measurement of position would be able to distinguish tunnelling and OBI, since only OBI would be detectable. 
%	\begin{itemize}
%		\item But this would require a detector positioned at the barrier, which is unfeasible in praxis (though there have been some theoretical studies using virtual detectors), since detectors are based on a time-of-flight measurements and are placed a significant distance away from the interaction region. 
%		\item An under-the-barrier transmission is a probability flux, which a hypothetical detector would be able to pick up on. 
%		\item even in tunnelling phenomenon other than strong-field physics, such as in a tunnelling junction, a position-resolved measurement would measure the probability density anyway. 
%	\end{itemize}
	An audience member suggests that localised position measurements would be able to distinguish tunnelling and OBI, since only OBI would be detectable.
	
	This gedanken experiment however would require a detector positioned at the atomic potential barrier, which is unfeasible in any kind of experimental setup since detectors require some time-of-flight information and are placed a significant distance away from the interaction region, of the order of several centimetres at least \cite{Doerner2000,Weger2013}. 
	There have been some theoretical studies using virtual detectors in combination with TDSE solutions \cite{Teeny2016,Teeny2016a}, but those again can not distinguish of course. 
	This is because both under-the-barrier and over-the-barrier transmission causes a probability flux of the wave function, which a hypothetical detector would be able to pick up without being able to distinguish between those two types of transmission.
	This remains the case also in different tunnelling scenarios such as in a tunnelling junction where a macroscopic, position resolving detector might be feasible.

	\item %representation of quantum wavefunction using trajectories: fundamentally: wave function doing something interesting. Wave function as a single wave packet (i.e. a gaussian), single trajectory mapping the peak of the wave packet represents that (expectation value) motion (i.e. group velocity approach). Weighted sum of trajectories representing a squewed wave packet. However, something that's not necessarily a wave packet, how can that be represented in a well-defined way by trajectories?
%	\begin{itemize}
%		\item wave function has a probability density, construct a distribution of trajectories with momenta and positions to represent that, using imaginary time and possibly complex space. Unless it is infinitely de-localized. 
%		\item Will an ensemble of (classical or quantum) trajectories not only represent an instantaneous probability distribution derived from a wave function, but also it's dynamics over time? Either quantum correction taken into account (i.e. proper quantum trajectories) or classical trajectories are only valid as long as Ehrenfest theorem is satisfied. 
%		\item This is exactly the domain of semiclassics and their methods, accounting for quantum non-locality, quantum corrections, and phases, resulting in the same qualitative description as the full quantum description. 
%	\end{itemize}
	The last point for this topic is concerning representation of a quantum wave function by using trajectories. 
	Fundamentally, we are trying to study the behaviour of a wave function doing something interesting. 
	In the most simple trajectory approach, the entire wave function is represented by a single simplified wave packet (i.e. a Gaussian) with the associated trajectory of its wave packet peak mapping the motion of the expectation value of the wave function (similar to a group velocity approach) \cite{heller_classical_1976}. 
	However, this approach can not describe a wave packet being split into a reflected and a transmitted part, and thus would either always remain bound or the bound state is fully depleted. 
	Trajectories representing a skewed wave packet \cite{Heller1975} would present a more generalised version. 
	% I couldn't really find anything better either, than Heller's (and further development by others...) thawed gaussian methods, but I think we can leave it at that?
	Even more accurate are descriptions that employ a large ensemble of trajectories following the probability distribution of the underlying wave function \cite{herman_semiclasical_1984}.
	
	The question is then: Will an ensemble of (classical or quantum) trajectories not only represent an instantaneous probability distribution derived from a wave function, but also its dynamics over time? This question was addressed in Ref.~\cite{Shvetsov-Shilovski2013} that studies the validity of the two-step semiclassical model disregarding quantum interference but accounting for the Coulomb field for strong-field ionisation. The Ehrenfest theorem \cite{Ehrenfest1927} (see, e.g., Ref.~\cite{Messiah1966} for a textbook treatment), which establishes quantum mechanical analogues of classical Hamiltonian equations, was applied in Ref.~\cite{Shvetsov-Shilovski2013}. Furthermore, the analysis of Ref.~\cite{Shvetsov-Shilovski2013} is based on a quantitative comparison of the electron momentum distributions obtained within the two-step model and by numerical solution of the TDSE. 
%	It was found that the discrepancy between the quantum result and the two-step model correlates with the situations where the ensemble average of the force acting on the electron from the ionic core deviates considerably from the same force calculated at the average position of the trajectories of the ensemble.  
%	It was found that the the two-step model differs from the quantum results when the ensemble average of the force acting on the electron from the ionic core deviates considerably from the same force calculated at the average position of the electrons in the ensemble.
	Reference \cite{Shvetsov-Shilovski2013} introduces the measure for the deviation of the dynamics of an ensemble of classical trajectories from the Ehrenfest’s theorem. This measure is the relative deviation between the force at the average position of the ensemble of trajectories and the average of the forces on the ensemble. A correlation was found between the invalidity of the two-step model and the deviation of the dynamics from the Ehrenfest's theorem.
	
	The general trends for the applicability of the two-step model in terms of laser intensity, wavelength, ellipticity, as well as in terms of the potential properties are identified in \cite{Shvetsov-Shilovski2013}. However, this study is done in the two-dimensional (2D) case and needs to be extended to the 3D one.
\end{itemize}

\section{Theoretical approaches to quantum tunnelling time}
\label{Sec:Numerical}

The guiding questions for the third and last topic are: 

\begin{quote}
	\emph{What are theoretical approaches used to investigate quantum tunnelling times? \\
		What are their various characteristics, advantages and disadvantages?}
\end{quote}

\subsection{Overview:}

For the overview, a brief and non-exhaustive list of different calculation approaches to the tunnelling time are given. They are categorised with regards to their theoretical foundation. 

\subsubsection{Quantum methods based on time-dependent Schrödinger equation}

First are numerical solutions to TDSE.
A common advantage of all these methods is that they are fully quantum calculations for the entire process. 
Further individual characteristics, advantages and disadvantages can be summarised as follows.

TDSE calculations which employ Coulomb vs Yukawa potentials \cite{Sainadh2019,Torlina2015} found that attoclock signal shows a prominent offset angle with Coulomb binding potential, while the offset angle vanishes for a Yukawa potential. This comparison offered an indirect proof of instantaneous tunnelling by comparing the results depending on the two different binding potential of the parent ion.

The numerical saddle-point method \cite{Eicke2018} uses a trajectory-free language and establishes a connection between the final momentum of the photoelectron and the numerical saddle-point time for the full Hamiltonian including the Coulomb potential. It supports the conclusion of instantaneous tunnelling. However, this method is gauge dependent.

The functional derivative method \cite{Ivanov2018} investigates the instantaneous ionisation probability as a functional derivative of the total ionisation with respect to the wave form of the ionising field, but does not map directly to any experimental observables. 
It is gauge independent, and found vanishing delay (or vanishing delay asymmetry with respect to the local peak in the field).

Bohmian mechanics \cite{Douguet2018} present a mapping from the quantum world to the trajectory language. 
However, the calculation is guided by a pilot wave not pertaining solely to the (eventually) ionised part of the wave packet near the tunnel exit, thus potentially giving false tunnelling information.
A separation of the (eventually) ionised part and bound part of the wave packet near the tunnel exit is, unfortunately, impossible, due to quantum nonlocality.

\subsubsection{Quantum methods based on strong-field approximation}

Strong-field-approximation (SFA) \cite{Keldysh1964,Faisal1973,Reiss1980} based quantum methods describe ionisation as a transition from an initial state unaffected by the laser field to a Volkov state, i.e., the free electron wave function in an electromagnetic field. 
Therefore, the SFA disregards the intermediate bound states and the ionic potential (e.g., Coulomb interaction) in the final state. 
Presently, several SFA-based quantum approaches are developed. 
Typically, these approaches decide which force dominates the trajectory of a photoelectron based on its position in space and use the corresponding approximations.
This separation and reduction of the acting forces allows for analytic calculations.
The imaginary part of the saddle-point time in SFA calculations relates to the inverse tunnelling rate, while the real part in these models is often taken as the tunnel exit time.

\begin{figure}[h]
	\centering
	\includegraphics[width=0.9\columnwidth]{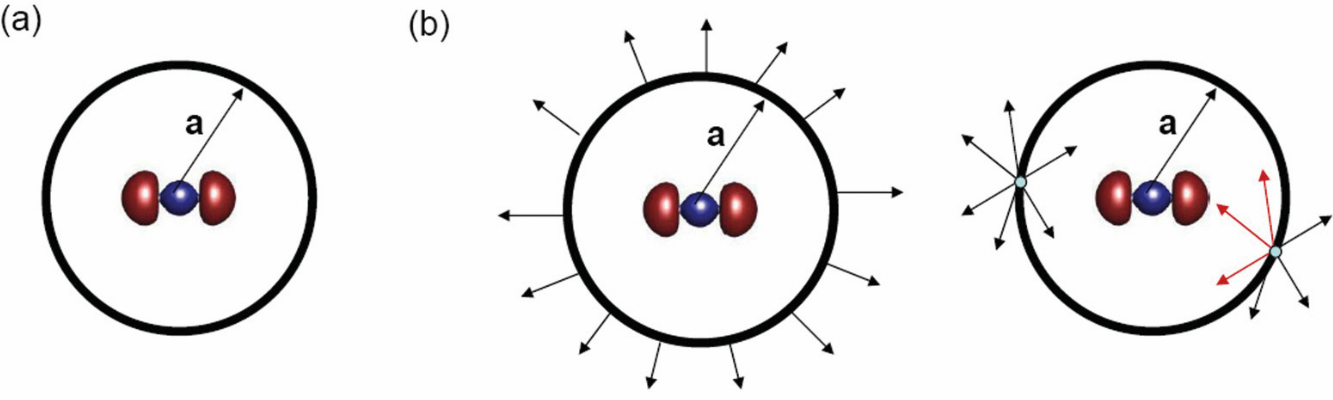}
	\caption{Separation of space into inner (close to parent ion) and outer (far away) regions of space. Adapted from \cite{Torlina2012}.}
	\label{fig:arm-sfaillustration}
\end{figure}

The analytic R-Matrix (ARM) method \cite{Torlina2012,Torlina2015} separates space into an inner region (Coulomb \& Laser field considered) and an outer region (Coulomb field neglected, eikonal-Volkov approximation), as illustrated in figure \ref{fig:arm-sfaillustration}.
The disadvantage of this method is the challenge of choosing proper integration contours for each trajectory.

\begin{figure}[h]
	\centering
	\includegraphics[width=0.7\columnwidth]{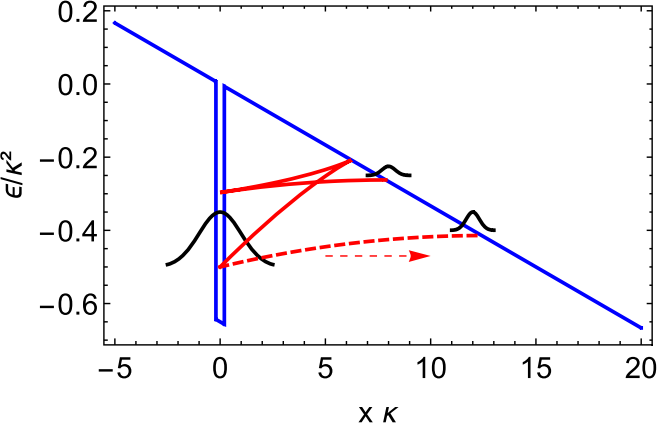}
	\caption{Illustration of rescattering and transmitting quantum trajectories under the potential barrier. Adapted from \cite{Klaiber2018}.}
	\label{fig:klaiberunderbarrier}
\end{figure}

The under-barrier recollision theory \cite{Klaiber2018} specifically includes interference between under-barrier rescattered and direct trajectories, as shown in figure \ref{fig:klaiberunderbarrier}.
This leads to a shift in the momentum wave packet peak, which can be interpreted as a delay. 
However, this method ignores Coulomb corrections.

\subsubsection{Hybrid quantum-classical method}

\begin{figure}[h]
	\centering
	\includegraphics[width=\columnwidth]{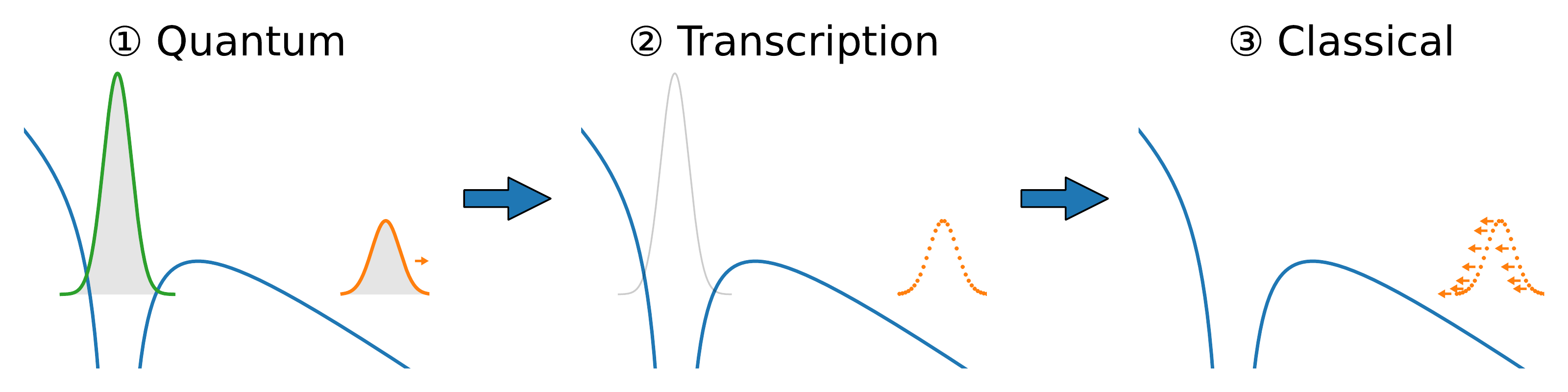}
	\caption{Concept of the backpropagation method.}
	\label{fig:ni-backpropagation}
\end{figure}

The backpropagation method \cite{Ni2016,Ni2018,Ni2018a} is a hybrid quantum-classical approach offering a unique perspective on the tunnelling process. 
It combines a fully quantum calculation of the ionisation process with forward propagation utilising TDSE solution, followed by a transcription of the resulting ionised quantum wave packet into classical trajectories, and a subsequent propagation of the trajectories backward in time, see figure \ref{fig:ni-backpropagation} for a sketch.
Another variant of the backpropagation method would be putting a sphere of virtual detectors \cite{Feuerstein2003,Wang2013,Wang2018,Xu2021,Wang2017,Zhang2017} around the target, where the flux is converted into classical trajectories during the laser pulse on the fly \cite{Liu2018,Ni2020}.

Why backpropagation? Firstly, as everyone agrees, tunnelling is a purely quantum process. Introducing a tunnelling barrier into the description of tunnelling ionisation, however, brings in clearly classical elements into the picture. Namely, tunnelling is now depicted with local tunnelling exit positions and momenta, which calls for a classical formulation.
Secondly, due to quantum nonlocality, the portion of the wave packet that would eventually be freed and the portion that would finally remain bound can not be separated during the tunnelling process. A separation is only possible in the far field, when these two portions are spatially detached.
These are exactly the design philosophy of the backpropagation method, a hybridisation of quantum forward and classical backward propagation.
It combines the advantages of the quantum and classical methods by offering the capability to include the full Hamiltonian and quantum tunnelling dynamics while retaining the local information from the classical trajectories.
It also naturally includes nonadiabatic tunnelling effects, automatically remove the offset angle from Coulomb effects, and retrieves the electron characteristics at the tunnel exit.

The classical backpropagating trajectories may be stopped whenever a certain condition is met, which defines the tunnel exit, yielding highly differential information of the tunnel exit.
In this manner, the backpropagation method may act as a common ground to compare different definitions of tunnelling.
It was found that a vanishing tunnelling time results if the tunnel exit is defined in the momentum space when the velocity of the trajectory vanishes in the instantaneous field direction (the velocity criterion), while defining the tunnel exit as a certain position in the coordinate space (the position criterion) gives rise to a finite tunnelling time \cite{Ni2018,Ni2018a}.
Different definitions of the tunnel exit were thus believed to be the origin of the tunnelling time debate.
It was further argued that the position criterion leads to inconsistencies and difficulties and thus the velocity criterion is favoured as the definition of the tunnel exit, and the tunnelling time delay should thus vanish \cite{Ni2018,Ni2018a}.

The backpropagation method has further enabled a study of the tunnelling time delay induced by orbital deformation \cite{Liu2018} and a subcycle time resolution of the linear laser momentum transfer, where a coupling between the nondipole and nonadibatic tunnelling effects was found \cite{Ni2020}.

\subsubsection{Semiclassical methods}

Semiclassical methods apply classical trajectories to describe the motion of an electron after it has been released from an atom or molecule by the laser pulse. The two-step \cite{vanLinden1988,Gallagher1988,Corkum1989} and the three-step \cite{Kulander1993,Corkum1994} models are the most widely known examples of the semiclassical approaches. These models do not account for the effect of the ionic potential on the electron motion in the continuum. Presently there are many trajectory-based models that do account for the ionic potential in the classical equations of motion. Among these are: Trajectory-based Coulomb SFA (TCSFA) \cite{Yan2010,Yan2012}, Quantum trajectory Monte-Carlo method (QTMC) \cite{Li2014a}, Coulomb quantum orbit strong-field approximation (CQSFA) \cite{Faria2015,Faria2017a,Faria2017b,Faria2018a,Faria2018b,Faria2018c}, semiclassical two-step model (SCTS) \cite{Shvetsov-Shilovski2016}, Quasistatic Wigner method \cite{Camus2017}, etc. The three-step model using complex classical trajectories \cite{Koch2020} and the classical Keldysh-Rutherford model \cite{Bray2018b} are closely related to this group of models.

Using a purely classical description of the electron motion it is not possible to describe the quantum interference effect in the photoelectron momentum distributions and energy spectra. Recently substantial progress has been achieved along these lines.  Along with some other approaches, the TCSFA, QTMC, CQSFA, and SCTS models account for interference effects. In these approaches every classical trajectory is assigned to a certain phase, and the contributions of different trajectories leading to a given final electron momentum are added coherently.

The TCSFA extends the well-known Coulomb-corrected strong-field approximation (CCSFA) ~\cite{Popruzhenko2008a,Popruzhenko2008b} by treating the laser field and the Coulomb force acting on the electron from the ion on an equal footing. The TCSFA accounts for the Coulomb potential in the phase of every trajectory within the semiclassical perturbation theory. The same approach is used in the QTMC model. In contrast to this, the SCTS and the CQSFA models account for the Coulomb potential beyond the semiclassical perturbation theory.

The quasistatic Wigner method \cite{Camus2017} employs the concept of the dominant quantum path. Using the space-time propagator, the quasistatic Wigner method considers the propagation of the electron wave function that originates from the initial bound state in the classically forbidden domain. The quasistatic description of the laser field is used in Ref.~\cite{Camus2017}. The phase of the quantum mechanical propagator determines the most dominant path along the tunnel channel, and therefore, determines the Wigner trajectory. The Wigner trajectory is merged with the corresponding classical trajectory in the continuum, see Ref.~\cite{Camus2017}. In this way the quasistatic Wigner method determines the initial conditions for the classical trajectory. It should be emphasised that the initial conditions include not only an initial momentum, but also a time delay. However, this method reduces the wave packet to a single trajectory. It should also be noted that the Wigner time is ill-defined in the tunnelling process \cite{Hofmann2019}. 

\begin{figure}
	\centering
	\includegraphics[width=0.7\columnwidth]{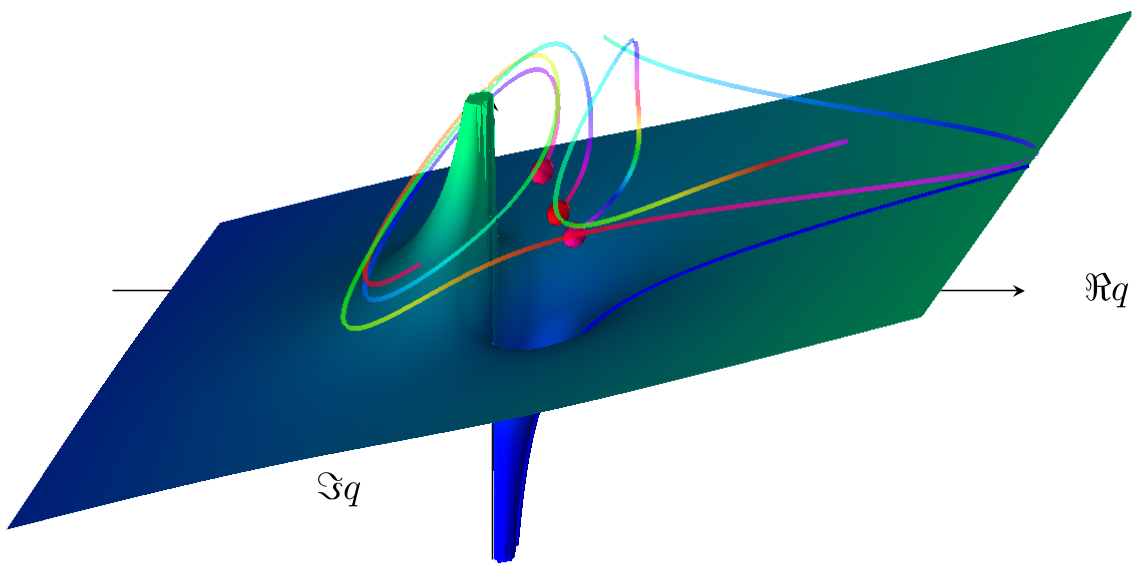}
	\caption{Schematic depiction of trajectories in complex space surrounding the singularity of the 1D radial Coulomb potential. Note the repulsive nature of the potential in the negative half of the plane. Orbiting trajectories result from propagation in complex time prior to ionisation \cite{Koch2020}.}
	\label{fig:complexintroschematic}
\end{figure}

Since real-valued trajectories are not able to describe tunnel ionisation, the complex-time-and-space model \cite{Koch2020} employs complex trajectories, as illustrated in figure~\ref{fig:complexintroschematic}. 
This approach was applied to the HHG process in Ref.~\cite{Koch2020}. All components of the three-step model are described in Ref.~\cite{Koch2020} within a single consistent trajectory framework. The trajectories are sampled from an initial Coulomb eigenstate, and the time propagation is performed using the final value coherent state propagator (see Ref.~\cite{Zamstein2014}). As a result, the model provides a unified and seamless trajectory description of the ground state, tunnelling, and collision process. The model shows quantitative agreement with fully quantum results. However, the contour in the plane of complex time, which is necessary to implement the model, has to be chosen manually.
%\rot{include figure(s) from Werner's "complexintro" slides showing preliminary tunnelling time results? What is the difference between the two figures?} 

\subsubsection{Classical methods}

And finally, purely classical models are still also developed and used often.
The Keldysh-Rutherford model \cite{Bray2018b} applies the famous Rutherford scattering formula taking the vector potential of the laser pulse as the asymptotic electron velocity and the Keldysh tunnelling width as the impact parameter. The model was tested by comparison of its predictions with the numerical solution of the TDSE using the hydrogenic potential and the screened (Yukawa) potential. In the latter case the action of the Coulomb field was gradually switched off. The striking similarity between the attoclock offset angle and the Rutherford scattering angle was revealed in Ref.~\cite{Bray2018b}. The Keldysh-Rutherford model suggests that the offset angle has a largely Coulombic origin \cite{Bray2018b}. Therefore, the model is questioning the interpretation of this angle in terms of a finite tunnelling time. However, the Keldysh-Rutherford model completely neglects nonadiabatic effects, and is also limited in its validity to short pulse durations and (relatively) weak intensities which are outside the typical parameter range of experiments to date. Therefore, some further work along this direction is needed.

Classical trajectory Monte Carlo (CTMC) methods \cite{Pfeiffer2012,Landsman2014b} are the classical cousin of QTMC, and often employed where interference effects are not of any key interest.
Since the calculations are computationally cheap compared to TDSE solutions and fewer trajectories are needed than in QTMC to reach similar statistical quality, these methods are able to fully include the ion Coulomb potential together with the laser field during the propagation after the tunnel exit, as well as various non-adiabatic effects \cite{Hofmann2014}, certain multi-electron effects \cite{Emmanouilidou2015} including Stark shift and an induced dipole in the parent ion \cite{Pfeiffer2012}.

\subsection{Debate:}
\begin{itemize}
	\item %regarding complex-time-complex-space method, how was the contour chosen manually?
%		\begin{itemize}
%			\item every time a trajectory orbits the singularity at the nucleus, there is a possibility for the quantum trajectory to be emitted from the bound wave packet and leave as part of the ionisation wave packet. Exact choice for after how many orbits an ionisation event happens is done by comparing to the full quantum result, for sections of initial coordinate space. Observables are then computed from the resulting trajectories. 
%			\item So the agreement to quantum calculations (TDSE) is by construction?
%			\item It is a discrete choice and not a fully tunable parameter. While there is a choice to match the quantum result, each discrete choice results in significantly different results, and it is not a fitting process. Interpretation of this choice is not clear yet from a physical point of view. 
%			\item can eventually bound from eventually ionised part of the wave packet be separated, and tunnelled part be traced through potential barrier?
%			\item trajectories far from core in long time limit are considered ionised. Searching for conditions (zero momentum for example) along those trajectories yields two complex times, labelled as tunnel entry and exit, where the difference can be interpreted as tunnelling time. Or to compare to experiments, difference to field maximum.
%		\end{itemize}
	Regarding the complex-time-and-space method \cite{Koch2020}, the question is raised how exactly the integration contour is chosen manually. 
	Every time a trajectory orbits the singularity at the nucleus, there  is a possibility for the quantum trajectory to be emitted from the bound wave packet and leave as part of the ionisation wave packet. 
	The exact choice for after how many orbits an ionisation event happens is done by comparing to the full quantum result, for sections of initial coordinate space. 
	Observables are then computed from the resulting trajectories.
	The number of loops is a discrete choice and not a fully tunable parameter.
	While there is a choice to match the quantum result, each discrete choice yields significantly different results, so the agreement with TDSE calculations is not entirely by construction.
	The interpretation of this choice is not clear yet from a physical point of view. 
	
	When it comes to separating the eventually bound from the eventually ionised part of the wave packet, trajectories far from the core in the long time limit are considered ionised.
	Searching for conditions (zero momentum for example) along those eventually ionised trajectories yields two complex times, labelled as tunnel entry and exit, where the difference can be interpreted as tunnelling time, see figure \ref{fig:complexintrotimesA}.
	Alternatively, the difference to the field maximum can be computed, as shown in figure \ref{fig:complexintrotimesB}. This may be required to compare the results to experiments where tunnel entry times may not be accessible but it does ignore a significant contribution to the total tunnelling time. As Fig.~\ref{fig:complexintrotimes} shows, the time required for tunnelling is non-zero in both real and imaginary components. Furthermore, a single averaged result may be insufficient to characterise the tunnelling process as the distribution of times is wide and asymmetric.
	\begin{figure}[h]
		\centering
		\subfigure[$\tau = t_{\mathrm{exit}}-t_{\mathrm{entry}}$]{\includegraphics[width=\columnwidth]{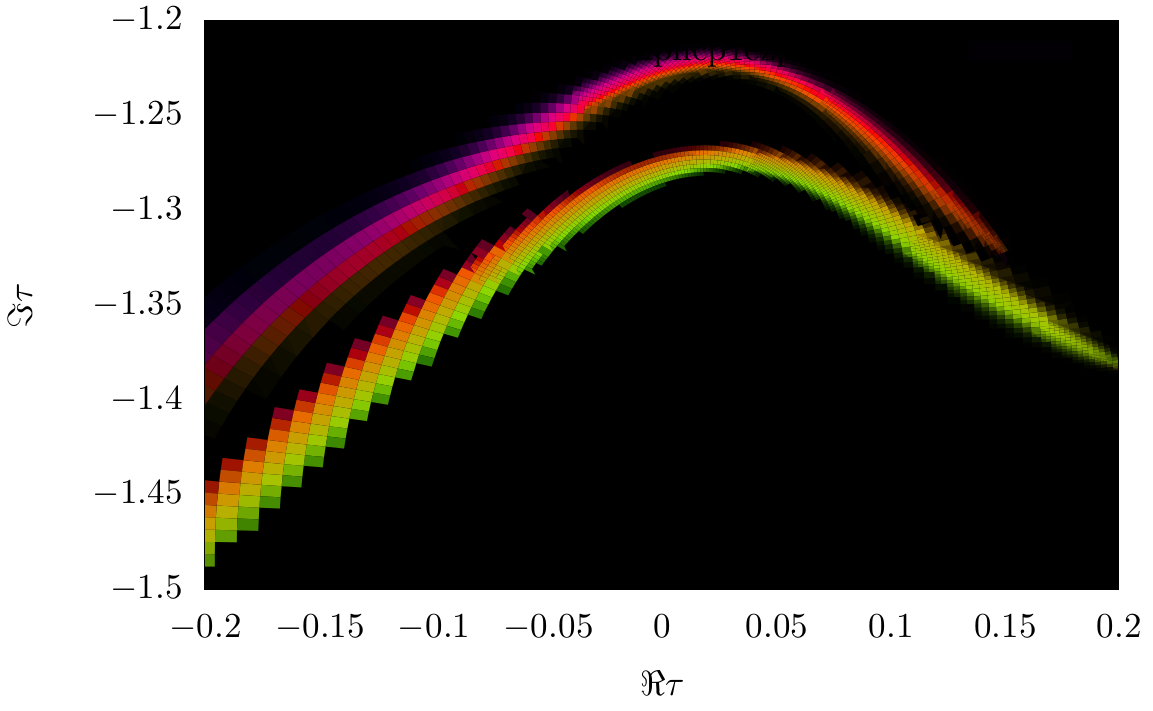} \label{fig:complexintrotimesA}}
		\subfigure[$\tau = t_{\mathrm{exit}}-t_{\mathrm{peak}}$]{\includegraphics[width=\columnwidth]{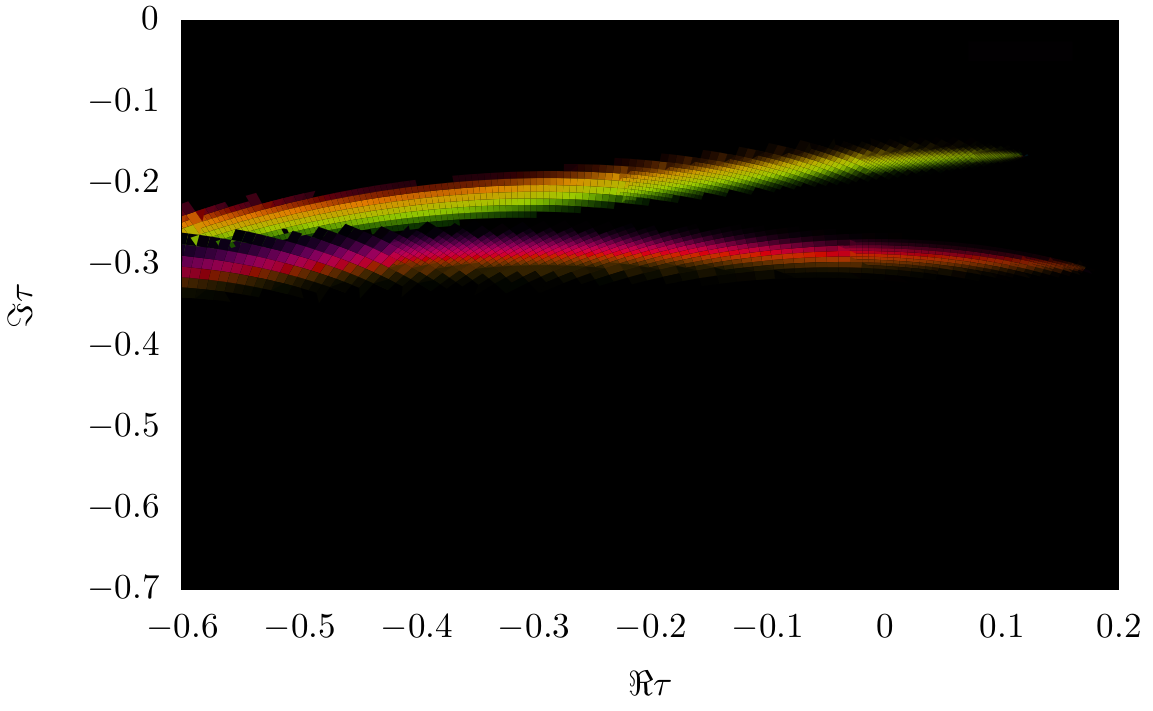} \label{fig:complexintrotimesB}}
		\caption{Tunnelling time distribution extracted from the complex-time-and-space method for a half cycle pulse, with wavelength $\lambda = 1033~\mathrm{nm}$, ionisation potential $I_p = 13.6~\mathrm{eV}$ (hydrogen), intensity $I = 1.9 \times 10^{14}~\mathrm{W/cm^2}$, resulting in a Keldysh parameter of $\gamma = 0.6$. Brightness encodes probability magnitude and colour encodes phase of the trajectory. Two distinct distributions belonging to two separate classical processes in the trajectory ensemble are visible in both plots.}
		\label{fig:complexintrotimes}
	\end{figure}
	Two distinct classical processes are found, and the two references (entry point or field maximum) for the tunnelling delay time differ significantly.
	
	\item % Question: effect of excited state in attoclock measurements
%		\begin{itemize}
%			\item was dealt with in \cite{Sainadh2019} study. H 1s vs 2s as bound state result in completely different final momentum distribution (2s leads to much smaller absolute momenta, different structure, and is less likely to happen). Contributions from different initial states can be separated.
%		\end{itemize}
	An audience member asks what the effect of excited states on an attoclock measurement is.
	This was dealt with in the \cite{Sainadh2019} study since molecular hydrogen had to be split into atomic hydrogen.
	In their extended data figures \& tables, it is shown how initial bound states 1s or 2s result in completely different final momentum distributions.
	Photoelectrons ionised from 2s have much smaller absolute momenta, their distribution shows a different structure, and the event is less likely to happen. 
	Therefore, contributions from different initial states can be separated.

	\item %SCTS method: tunnelling vs over-the-barrier ionisation distinction?
	
	Already in section \ref{Sec:Philosophical}, combined approaches which offer quantum behaviour with trajectory insight have been identified as beneficial for many strong-field (tunnelling) phenomena models. 
		
	Typically, the SCTS model requires large ensembles of classical trajectories to resolve fine interference details. These trajectories are propagated, and their final momenta are binned in cells in momentum space. This is often referred to as ``shooting method" \cite{Yan2010}, although this approach has nothing to do with the shooting method for solving a boundary value problem. In contrast to the TCSFA, QTMC, and SCTS, the CQSFA method finds all the trajectories corresponding to the given final momentum. This approach is often called the solution of the ``inverse problem" and it allows to bypass the necessity of large ensembles of trajectories. However, the solution of the inverse problem is a non-trivial task, and, furthermore, is generally less versatile than the ``shooting method".
	
	Any trajectory-based model requires specification of initial conditions, i.e., the initial electron velocity and the starting point of the classical trajectory. Indeed, these initial conditions are needed to integrate the Newton's equations of motion.  The starting point, i.e., the tunnel exit, is found using the separation of the tunnelling problem in parabolic coordinates \cite{Landau1965}. The Stark shift of the energy level that has an effect on both the tunnel exit and ionisation probability was also taken into account in the SCTS. It is generally considered in the semiclassical models that the electron departs with zero initial velocity along the laser polarisation direction $v_{0,\parallel}=0$ and an arbitrary initial velocity $v_{0,\perp}$ in the perpendicular direction. The ionisation times and the initial transverse velocities are distributed in accord with the static ionisation rate:
	\begin{equation}
	\label{tunrate}
	w\left(t_{0},v_{0, \perp}\right)\sim\exp\left(-\frac{2\kappa^3}{3F\left(t_0\right)}\right)\exp\left(-\frac{\kappa v_{0,\perp}^{2}}{F\left(t_0\right)}\right)
	\end{equation}
	with $\kappa=\sqrt{2I_{p}}$. The quasistatic approximation is used in Eq.~(\ref{tunrate}), i.e., the static field strength $F$ is replaced by the instantaneous value $F\left(t_0\right)$. The quasistatic approximation is used in both QTMC and the SCTS.
	
	We note that many trajectory-based models use the SFA formulas instead of Eq.~(\ref{tunrate}) to distribute the initial conditions of classical trajectories, see, e.g., Refs.~\cite{Yan2010,Boge2013,Hofmann2014,Geng2014,Brennecke2020}. This allows to investigate nonadiabatic effects in above-threshold ionisation and often leads to a better agreement with the numerical solution of the TDSE.
	%However, the validity of the SFA-based formulas as distributions of the initial conditions needs to be studied in a systematic way.
	Recently, the SFA-based formulas as distributions of the initial conditions have been validated in a systematic way \cite{Ma2021}. It is found that a combination of SFA initial conditions with complex weight and a trajectory model of SCTS provides the best solution for obtaining the most accurate attoclock signal \cite{Ma2021}.
	The SCTS model has not been extended to the over-the-barrier ionisation (barrier-suppression regime) yet. Such an extension can be easily done as discussed above, see Eq.~(\ref{parvel}).
	
	Recently an efficient extension and modification of the SCTS model was proposed \cite{Brennecke2020}. In its original formulation the SCTS model uses the phase of the semiclassical matrix element \cite{Miller1974,Walser2003,Spanner2003} (see Refs.~\cite{TannorBook2007,GrossmannBook2008} for a textbook treatment), but completely disregards the pre-exponential factor of the bound-continuum transition matrix element. The influence of this pre-exponential factor was for the first time studied in Ref.~\cite{Brennecke2020}. The modulus of the pre-exponential factor corresponds to the mapping from initial conditions for electron trajectories to the components of the final momentum. It affects the weights of classical trajectories. The phase of the pre-exponential factor modifies the interference structures. This phase is known as a Maslov phase and can be viewed as a case of Gouy’s phase anomaly, see Ref.~\cite{Brennecke2020}. Furthermore, a novel approach to the inverse problem applying a clustering algorithm was proposed in \cite{Brennecke2020}. The modified version of the SCTS demonstrates excellent agreement with numerical solution of the TDSE for both photoelectron momentum distributions and energy spectra. It was found that the account for the pre-exponential factor is crucial for the quantitative agreement with the TDSE. This novel version of the SCTS can be applied not only to linearly polarised laser fields, but also to non-cylindrically-symmetric ones, e.g., bicircular laser pulses \cite{Brennecke2020}.
	
	The recent semiclassical two-step model with quantum input (SCTSQI) \cite{Shvetsov-Shilovski2019a} is a mixed quantum-classical approach that combines the SCTS with the numerical solution of the TDSE. To perform the synthesis of the trajectory-based approach with the TDSE, the Gabor transformation of the wave function $\Psi\left(x,t\right)$
	\begin{eqnarray}
	\label{Gabor}
	& &G\left(x_0,p_x,t\right)=\frac{1}{\sqrt{2\pi}}\int_{-\infty}^{\infty}{\Psi}\left(x^{\prime},t\right)\exp\left[-\frac{\left(x^{\prime}-x_{0}\right)^2}{2\delta_{0}^{2}}\right]\nonumber\\
	&\times&\exp\left(-ip_xx^{\prime}\right)dx^{\prime},
	\end{eqnarray}
	was used in the SCTSQI \cite{Shvetsov-Shilovski2019a}. Here $x_0$ is the point in the vicinity of which the Gabor transform is calculated and $\exp\left[-\frac{\left(x'-x_0\right)^2}{2\delta_{0}^{2}}\right]$ is a Gaussian window of the width $\delta_{0}$. The quantity $\left|G\left(x_0,p_x,t\right)\right|^2$ describes the momentum distribution of the electron near the point $x_0$ at time $t$. This is nothing just the Husimi distribution, which can be also obtained by Gaussian smoothing of the Wigner function.
	
	In Ref.~\cite{Shvetsov-Shilovski2019a} the Gabor transform (\ref{Gabor}) was used in combination with the absorbing boundaries that prevent the unphysical reflections of the wave function from the grid boundary. More specifically, the Gabor transform was applied to the part of the wave function that is absorbed at every time step of the solution of the TDSE. Figure \ref{fig:nikolay_husimi} shows an example of the corresponding Husimi distribution calculated at the end of a few-cycle laser pulse. This absorbed part is transformed in the ensemble of classical trajectories that is propagated using classical equations of motion. Therefore, initial positions and momenta of classical trajectories used to simulate an electron wave packet are extracted from the exact quantum dynamics. It is clear that the convergence with respect to the position of the absorbing boundaries and the number of trajectories launched at every time step should be checked in this approach. The absorbing boundaries must be far enough to not affect the bound part of the wave function. The SCTSQI yields quantitative agreement with quantum results \cite{Shvetsov-Shilovski2019a}. What is even more important, it corrects the inaccuracies of the standard trajectory-based approaches in description of the ionisation step and circumvents the complicated problem of choosing the initial conditions.
	
	\begin{figure}[h]
	\begin{center}
	\includegraphics[width=0.5\textwidth]{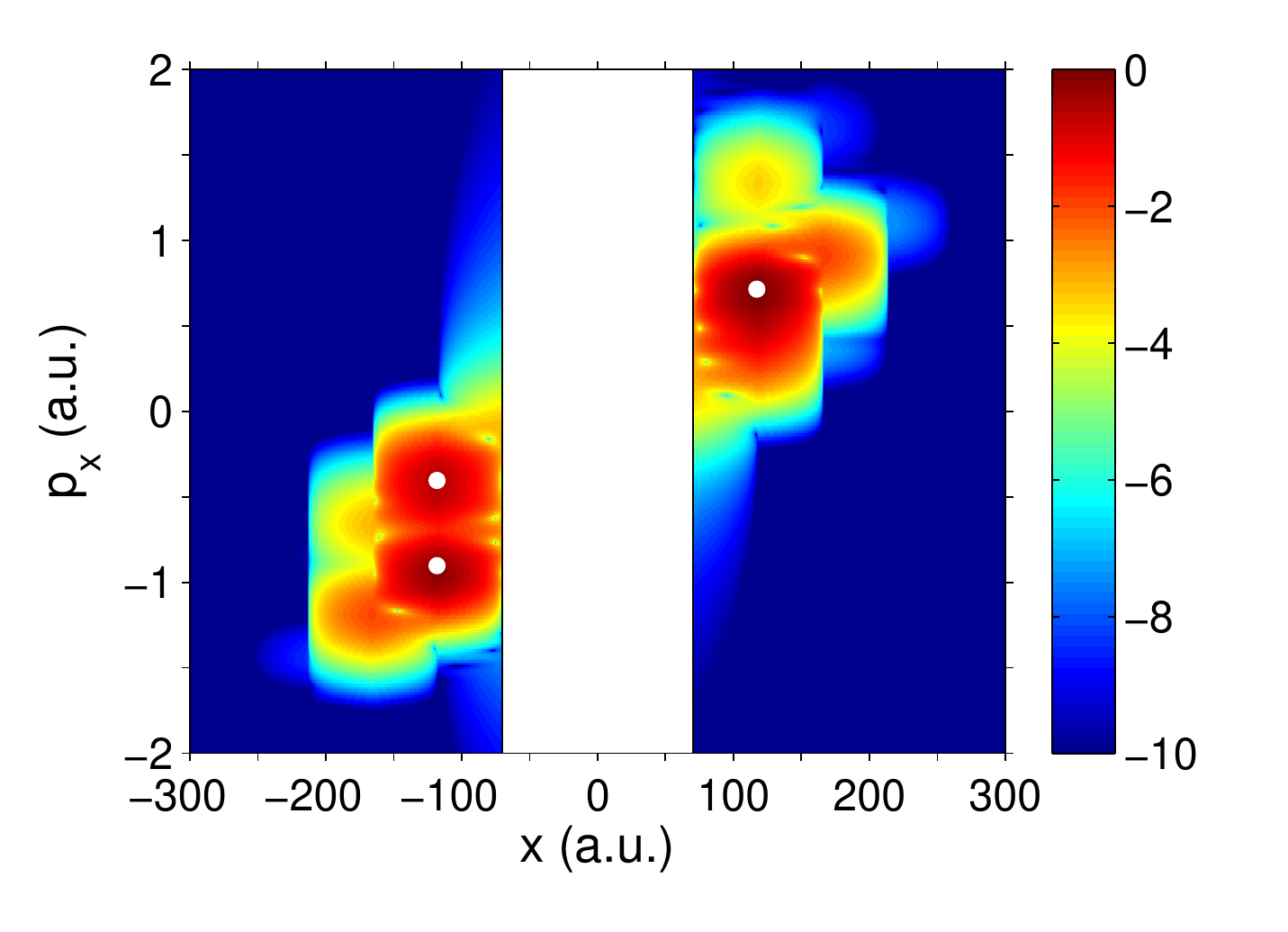} 
	\end{center}
	\caption{The Husimi distribution $\left|G\left(x_,p_x,t\right)\right|^2$ in the absorbing mask regions calculated for ionisation of 1D model atom at the end of the laser pulse with a duration of 4 optical cycles, intensity of $2.0\times10^{14}$ W/cm$^2$, and a wavelength of 800~nm. A logarithmic colour scale is used. The three main maxima of the Husimi distribution are shown by white circles.}
	\label{fig:nikolay_husimi}
	\end{figure}
	
	However, future work is needed to turn the SCTSQI model in a powerful tool for studies of tunnelling. First, the model formulated for the one-dimensional (1D) model atom should be generalised to the three-dimensional case (3D). To describe fine details of interference patterns accurately enough, large numbers of classical trajectories are needed in the SCTSQI. In addition to this, the ensembles of trajectories are launched at every step of the time propagation. As the result, the SCTSQI model includes all possible trajectories, and it is not always easy to distinguish between them. This hampers the understanding of the strong-field phenomena that is expected to be provided by the SCTSQI model and its future extensions.
	Therefore, the number of trajectories has to be reduced in the SCTSQI approach, e.g., by using more sophisticated sampling techniques.
	
	%		\begin{itemize}
	%			\item Over-the-barrier is not included in the SCTS model. Initial conditions required for semiclassical approaches (input). Usually zero momentum along tunnelling direction is assumed with arbitrary perpendicular momentum.  Often quasistatic formulas are used to compute distribution of instances of ionisation, and tunnel exit point. For over-the-barrier cases, different initial conditions would need to be defined, for example initial position at top of suppressed barrier, with velocity based on conservation of total energy from the ground state. 
	%			\item newer SCTS with quantum input (SCTSQI) \cite{Shvetsov-Shilovski2019a}, synthesis of TDSE and semiclassical model using Husimi distribution. This means that all possible trajectories are included in principle, but they can not easily be distinguished, and the work on this model is ongoing. 
	%		\end{itemize}
	%	\item SCTS method: distribution of initial conditions leading to ionisation in phase space: single area or multiple spots?
	%		\begin{itemize}
	%			\item SCTSQI approach, quasiprobability distribution from Gabor transform finds blobs of initial conditions in phase space, at a given time. 
	%		\end{itemize}
	\item %in SCTSQI method: mask function absorbing wave function over spatial extension. Probability of total wave packet degreasing?
%		\begin{itemize}
%			\item TDSE solved with absorbing boundaries, masking ionised part of wave function, Husimi distribution transforms absorbed part into ensemble of classical trajectories. Absorbing boundary should be far enough to not absorb bound part. Position-dependence for absorbing boundary must be checked as convergence test. 
%			\item issue remaining: wave function modulus (bound and ionised part) summed up is no longer 1 due to smooth mask function.
%			\item with a good mask, the sum should be close to unity, even if not exactly equal. mask should not be too close to core, and sum must be monitored, to ensure it remains reasonably close to 1.
%			\item efficacy of this depends on ionisation probability, how much of the wave function is going to hit the absorbing boundary. norms are generally not conserved with absorbing boundaries. 
%		\end{itemize}
	A mask function which is absorbing the wave function over a spatial extension, such as in the SCTSQI method for example, will lead to a decreasing total probability of the wave packet.
	This must be monitored over the course of the calculation to ensure it does not introduce unwanted artefacts through the choice of position or steepness of the absorbing mask.
	The efficiency of this also depends on the ionisation probability which determines how much of the wave function is going to hit the absorbing boundary. 
\end{itemize}

\section{Outlook}
It is evident that much remains to be done to further improve our general understanding of the tunnelling process as well as the interaction between the strong laser light and the target atom (or molecule, surface, liquid, \ldots) in order to tackle the underlying reasons for why so many approaches reach opposing conclusions.
Given the lack of a clear, agreed upon definition of the onset and conclusion of tunnelling, it is perhaps unsurprising that there is also not a clear pattern between classical or quantum methods in their various predictions regarding instantaneous or finite tunnelling time, let alone numerical values.
More than anything, this debate has demonstrated the need to find a common ground on which to compare the vast range of theoretical approaches and experimental setups.
One of the few prevailing themes of this debate that most everyone could agree on is that a combination of classical and quantum theory is required for describing tunnelling processes in order to be able to interpret the experimental evidence.

\begin{acknowledgement}
 C.H. acknowledges support by a Swiss National Science Foundation mobility fellowship. 
 W.K. has received financial support from the Israel Science Foundation (1094/16) and the German-Israeli Foundation for Scientific Research and Development.
 H.N. was supported by Project No. 11904103 of the National Natural Science Foundation of China (NSFC), Project No. M2692 of the Austrian Science Fund (FWF), Projects No. 21ZR1420100 and 19JC1412200 of the Science and Technology Commission
 of Shanghai Municipality, and the Fundamental Research Funds for the Central Universities.
 N.I.S. was supported by the Deutsche Forschungsgemeinschaft (Grant No. SH 1145/1-2).
\end{acknowledgement}

% The section below may be edited at your convenience to acknowledge 
% each author's contribution to the manuscript.
% You may remove it if you are a single author.
\section{Authors contributions}
All authors were involved in the live debate and the preparation of the manuscript.
All authors have read and approved the final manuscript.

\vspace{0.3cm}
This is a post-peer-review, pre-copyedit version of an article
published in The European Physical Journal D. The final authenticated version is available online at:
 \url{https://doi.org/10.1140/epjd/s10053-021-00224-2}.
%
% BibTeX users please use
% \bibliographystyle{epj.bst}
\bibliographystyle{epj}
\bibliography{references-2}

\end{document}